# Planar Laser-Induced Fluorescence system for Space and Phase-resolved Ion Velocity Distribution Function Measurements

S.H. Son[1,2], I. Romadanov[1], N.S. Chopra[3], and Y. Raitses[1]

[1] Department of Astrophysical Sciences, Princeton University, Princeton, NJ, USA.

[2] Princeton Plasma Physics Laboratory, Princeton, NJ, USA.

[3] Applied Materials Inc., Glouster, MA, USA.

**Abstract:**

In this work, we present a planar laser-induced fluorescence (PLIF) system for two-dimensional (2D) spatial and phase-resolved ion velocity distribution function (IVDF) measurements. A continuous-wave tunable diode laser produces a laser sheet that irradiates the plasma, and the resulting fluorescence is captured by an intensified CCD (ICCD) camera. Fluorescence images recorded at varying laser wavelengths are converted into 2D IVDFs using the Doppler shift principle. Comparing six image filters, the singular-value decomposition (SVD)-based noise filtering is identified as the most effective for enhancing the signal-to-noise ratio while preserving the IVDF structure. The developed ICCD-based PLIF system is tested in an electron-beam generated E × B plasma with a moderate bulk plasma density of $\sim 10^{10}$ cm$^{-3}$. The PLIF measurements are validated against a conventional single-point LIF method using photomultiplier tube (PMT)-based detection at various positions. The phase-resolving capability of the system is tested by oscillating the plasma between two nominal operating modes with different density profiles and triggering the ICCD camera with the externally driven plasma oscillation. The resulting oscillations in fluorescence intensity show good agreement with plasma density variations measured by electrostatic probes, demonstrating the system's ability to resolve phase-dependent dynamics. The measured IVDFs reveal several signatures of ion dynamics in this plasma source, including radially outflowing ions and anomalous ion heating in the plasma periphery, as anticipated by theoretical studies.

# Table of Contents

# 1. Introduction

Spatially resolved measurements of ion velocity distribution function (IVDF) is critical for low-temperature plasma physics and its applications as they provide detailed insights into transport, sheath dynamics, and other fundamental plasma processes [1–6]. Recently, the demand for temporal or phase-resolved measurements has also increased, in response to advances in modern plasma sources that exhibit various time-dependent characteristics of ion dynamics [7–9]. For example, E × B plasmas, including Hall thrusters, Penning discharges, and magnetic nozzles, feature various instabilities that lead to anomalous transport of ions [10–16]. Additionally, processing plasmas such as capacitively coupled plasmas (CCPs) and inductively coupled plasmas (ICPs) are frequently operated in pulsed modes [17–21] or driven by arbitrary bias waveforms [22–24], resulting in IVDFs that evolve over time. Therefore, diagnostics capable of resolving IVDFs in both space and phase are of broad interest for understanding and improving modern plasma sources.

Laser-induced fluorescence (LIF) is a widely used, non-invasive diagnostic technique for studying ion and neutral dynamics in various low-temperature plasma sources [25–29]. In this method, a laser optically pumps the plasma, and the resulting fluorescence emission is analyzed. By sweeping the laser wavelength across an absorption line of a target ion metastable state, IVDFs can be measured using the Doppler shift principle. Ion metastable species with velocity $v$ in along the laser feels shift $\Delta \lambda = \lambda_0 v/c$ to the laser set as $\lambda_0$ at lab frame, where c is the speed of light. As a result, metastable ions with a specific velocity are excited at a corresponding laser wavelength. Therefore, by mapping the emission intensity as a function of laser wavelength, the IVDF can be reconstructed assuming the target metastable ions are in thermal equilibrium with all ions. A laser with a narrow linewidth is preferred, as the linewidth sets the upper limit on the achievable ion velocity resolution. Among various lasers, continuous-wave (CW) tunable diode laser (TDL) is commonly used, as they provide narrower linewidth compared to other alternatives such as pulsed or CW dye lasers within a similar cost range [30].

In practice, conventional LIF diagnostics involve directing a focused laser beam into the plasma and collecting the resulting fluorescence at a single point (0D) using a photodetector, such as a photomultiplier tube (PMT). For spatially resolved measurements, spatial scans can be performed by mounting the laser beam, the fluorescence collection optics [26], or the plasma source itself [31] on translational stages. However, this scanning approach is inherently slow, as it requires a full laser wavelength sweep at each spatial position. Consequently, the total measurement time becomes N times longer than a single-point measurement, where N is the number of spatial positions of interest. Additionally, this approach limits each spatial point to be recorded at a different time, making it unsuitable for plasma sources that drift or evolve, even over relatively long timescales on the order of minutes or hours. For example, plasmas generated by thermionically emitting cathodes, which exhibit thermal drift [32], or Hall thrusters, which undergo gradual changes due to acceleration channel erosion [33], require alternatives to conventional LIF systems.

To overcome the limitations of point-by-point LIF scanning, planar laser-induced fluorescence (PLIF) techniques could be considered. In a PLIF system, a laser sheet illuminates a broad region of the plasma, and a camera with a wide field of view (FOV) captures a two-dimensional (2D) fluorescence map in a single acquisition, eliminating the need for mechanical spatial scanning. However, expanding the laser beam into a sheet inherently reduces its intensity, which in turn diminishes the resulting fluorescence signal. To compensate for this, most prior PLIF implementations have employed high-power pulsed lasers [7,21,34–38]. However, the broader linewidth of the pulsed lasers limits the velocity resolution of the measured IVDFs.

Only a few studies have reported PLIF or camera-detection LIF system using CW lasers that are favorable resolving high-resolution IVDFs. Since CW lasers are generally weaker than pulsed lasers and therefore induce less fluorescence, which, when combined with metastable quenching, often results in signals falling below detection limits [39], it is important to investigate the limitations of CW laser-based PLIF for its broader applicability to various plasma sources. Lee et al. [40] measured time-averaged 1D IVDFs in the sheath of a filament-discharge plasma using a low-power (~15 mW) CW TDL and an ICCD camera for detection. However, the reduced intensity of the expanded laser sheet was insufficient for diagnosing 2D maps of IVDFs. Paul and Scime [41] reported the first 2D-resolved IVDFs using a moderate-power (~70 mW) tunable CW dye laser shaped into a laser sheet and a fast-frame camera detection. However, this work was conducted on the PHASMA (Phase Space Mapping) experiment [42] with plasma densities on the order of $10^{12}$–$10^{13}$ $cm^{-3}$, which is several hundred to thousand times denser than typical sheath or bulk plasma densities encountered in industrial plasma sources. In summary, despite of recent advances, the applicability of PLIF for resolving 2D IVDFs in various low or moderate density low-temperature plasmas is not yet been demonstrated, both for time-averaged and phase-resolved measurements.

In this paper, we present the development of a PLIF system utilizing a compact, low-power (~20 mW) CW TDL for laser launch and an ICCD camera for detection. The system is implemented in an electron-beam-generated E × B plasma with a moderate bulk plasma density $\sim 10^{10}\ cm^{-3}$; a regime more representative of industrial and research plasma conditions. Using this setup, we demonstrate space (2D) and phase-resolved measurements of IVDFs and discuss the ion dynamics in this plasma source.

The structure of this paper is as follows:

**Section 2** introduces the experimental setup, including the target plasma source used as a test bench. **Section 3** provides details on the development of the PLIF system, including optical components, the operating scheme, and a quantitative analysis of image denoising methods. **Section 4** presents and discusses the results of both time-averaged and phase-resolved IVDF measurements. First, the system is directly validated against the conventional LIF method. Then, the signatures of ion dynamics in the target plasma source are discussed. The applicability of the PLIF system for phase-resolved IVDF measurements is subsequently demonstrated. Conclusions are provided in **Section 5**.

## 2. Experimental setup

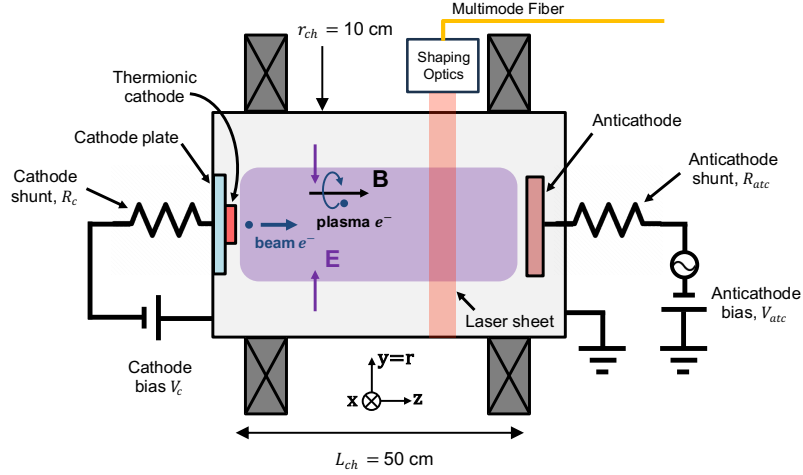

Figure 1. Schematic of the electron-beam generated E × B plasma source. See [43–47] for details.

An electron-beam generated E × B plasma source is used as a test bench for the developed diagnostics. A schematic of the plasma source is shown in Fig. 1(a).

This plasma source consists of a cylindrical vacuum chamber with a radius of $r_{ch} = 10\ cm$ and length of $L_{ch} = 50\ cm$, which is pumped down to a base pressure of 1 µTorr using a turbo-molecular pump. The chamber is filled with argon gas at a pressure of $0.1 - 0.5$ mTorr. An axial (z-direction) magnetic field of $50 - 100$ G is generated by a Helmholtz-like coil along the chamber. Electrons are thermionically emitted from a tungsten filament and accelerated across the cathode sheath in front of a stainless-steel plate with a radius of $r_c = 4.5\ cm$, located at one end of the chamber. The cathode and cathode plate are biased to a negative DC potential of $V_c = -55\ V$ relative to the grounded chamber. The electron beam is guided by the axial magnetic field, and a portion of the beam collides with neutral argon atoms, initiating a cylindrical plasma discharge. A laser sheet is aligned in the (r–z) plane of this plasma, as indicated by a red transparent box. In addition to the cathode plate, an electrically isolated, nonmagnetic stainless-steel plate, referred to as the anticathode, with a radius of $r_{atc} = 4.5\ cm$ is installed at the end of the chamber opposite the cathode. The chamber is equipped with various electrostatic probes that measure the radial profiles of plasma parameters, including a Langmuir probe (LP) to measure EEDF, floating emissive probe (EP) to measure plasma potential, and Ion probe (IP) to measure time-evolution of ion density under thick sheath assumption. More details on the electrostatic diagnostics could be found in [44].

We have investigated various characteristics of this plasma source under different operating conditions. For example, ion kinetics such as warm-ion production [45], electron kinetics[46], and instabilities such as rotating spokes [47] have been observed. Notably, by varying the voltage applied to the anticathode plate, the plasma can operate in one of two distinct modes—*Collector* mode and *Repeller* mode—as identified in our previous work [46]. In Collector mode, the anticathode is grounded ($V_{atc} = 0\ V$), allowing it to collect nearly all incident electrons. In contrast, in Repeller mode, the anticathode is biased to the cathode potential or lower ($V_{atc} \leq V_c$), repelling plasma and beam electrons back into the plasma volume. Consequently, Repeller mode produces a higher plasma density compared to Collector mode. This clear mode transition is particularly useful for evaluating the phase-resolving capability of the diagnostic system.

To evaluate the performance of the developed LIF system, this source is used as a test bench under two distinct operating scenarios, A and B, corresponding to (A) time-averaged and (B) phase-resolved measurements of spatially resolved IVDFs. The operating parameters for each case are summarized in Table 1. The radial profile of plasma parameters for scenario (A) is provided in [44,46], and that for scenario (B) is presented in **Section 4.2**. The electron temperature profile is not radially uniform due to the presence of the electron beam near the radial center. As a result of the source characteristics, the effective electron temperature is higher at the center and lower at the periphery (r = 20 mm).

In scenario (A), the anticathode is biased to a constant DC voltage of $V_{atc} = -55$ V. So, the plasma stays in *Repeller* mode. Note that a prominent azimuthal structure, rotating spoke, exists in the plasma bulk. In scenario (B), the anticathode is AC-modulated as $V_{atc} = V_0 \cos(2\pi ft)$, where $V_0 = -75$ V and f = 0.5 kHz. This modulation causes the plasma to oscillate slowly between *Collector* and *Repeller* modes with an oscillation period $\tau = 2$ ms, resulting in periodic fluctuations in plasma density. Phase-resolved LIF is employed to capture the corresponding variation in the fluorescence profiles associated with this density modulation. Note that the use of higher pressure and a weaker magnetic field in scenario (B) significantly suppresses the rotating spoke, making the Collector-to-Repeller mode transition the most prominent temporal feature.

This behavior is beneficial for diagnostic triggering, and it will be discussed further in the following section.

Table 1. Experimental condition of each operating scenarios, (A) for time-averaged measurements, and (B) for phase-resolved measurements. p: pressure, $V_c$: cathode voltage, $V_{atc}$: anticathode voltage or waveform, $B_z$: axial magnetic field intensity, $n_p$: plasma density measured at radial center ($r = 0\ cm$), $T_{e,0}$: effective electron temperature at center, $T_{e,2}$: effective electron temperature at the radial periphery ($r = 2\ cm$)

| | Scenario (A) for Time-Averaged | Scenario (B) for Phase-Resolved |
|---|---|---|
| $p\ [mT]$ | 0.1 | 0.5 |
| $V_c\ [V]$ | −55 | |
| $V_{atc}\ [V]$ | −55 | $V_0 \sin(2\pi f t)$ $V_0 = -75, f = 0.5\ kHz$ |
| $B_z\ [G]$ | 100 | 50 |
| $n_p\ [10^{10}\ cm^{-3}]$ | 2.5 | 2.5 − 4 (upon modulation) |
| $T_{e,0}\ [eV]$ | ~ 12 | ~ 5 |
| $T_{e,2}\ [eV]$ | ~ 3 | ~ 1.8 |

# 3. Diagnostics development

## 3.1. Optical Setup

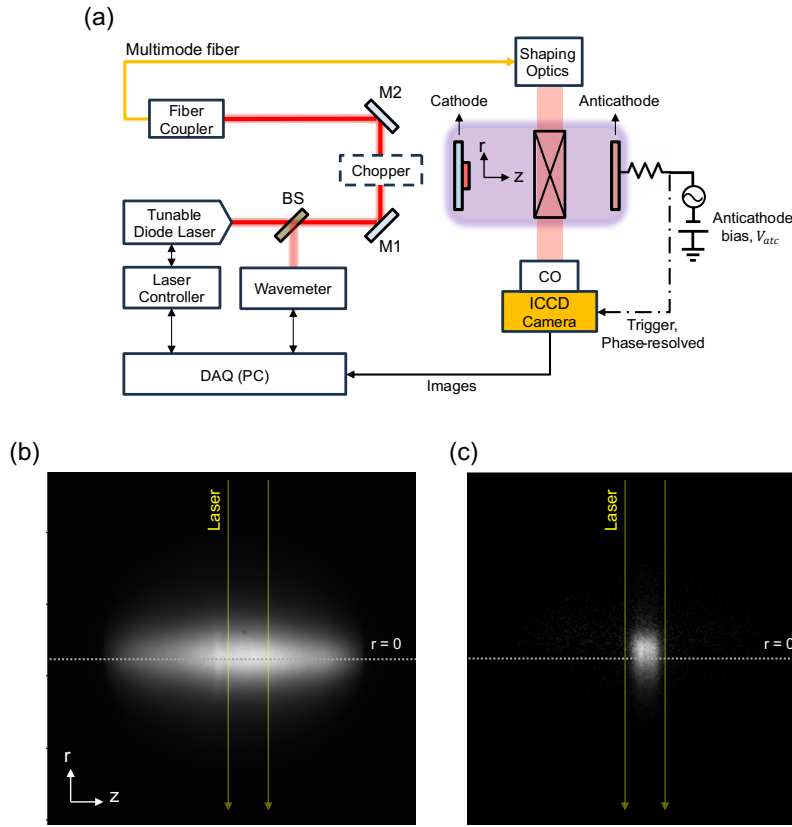

Figure 2. Schematic of the developed PLIF system. (a) Optical setup of the diagnostic system. The purple region indicates the plasma, and the black box with a cross mark indicates the region where laser-induced fluorescence (LIF) is produced. Optical components are labeled as follows: BS – beam splitter; M – mirror; CO – collection optics. (b) Emission pattern captured using a PI-MAX 4:1024i camera. (c) Fluorescence pattern extracted by subtracting the background from the (b). The plasma is operated under scenario (A), with the laser tuned to the central wavelength $\lambda = 668.6138\ nm$, within the wavemeter's uncertainty of $\Delta\lambda = \pm 0.0002\ nm$. The direction and edge of the incident laser sheet are indicated by yellow arrows. Both images are normalized to their respective maximum intensities for visualization purposes.

The schematic of the PLIF system developed in this study is shown in Fig. 2(a). The system is built around a tunable diode laser (TDL), specifically a single-mode TOptica DLC DL PRO 670, which features a Littrow-type grating-stabilized external cavity design. This TDL provides a coarse tuning range from 660 to 673 nm and a mode-hop-free tuning range of approximately 20 GHz. The output power is wavelength-dependent, with a maximum of approximately 23 mW. The laser wavelength is controlled by scanning the voltage applied to the laser's piezo actuator. A portion of the laser beam is sampled by a beam splitter (BS) and directed to a wavemeter (Bristol, 621-A), which features built-in continuous calibration using a single-frequency He–Ne laser and a specified accuracy of ± 0.0002 nm at 1000 nm. The wavemeter's reading accuracy exceeds the TDL's linewidth, which limits the achievable velocity resolution to approximately $\Delta v = 89$ m/s, as estimated from the Doppler shift principle. The corresponding resolution of the broadening, i.e. ion temperature is $\frac{1}{2} M_i (2\Delta v)^2 = 0.007$ eV.

Various transition schemes are known for argon ion metastable states [30], but careful selection is necessary, as the metastable population can vary depending on the characteristics of the electron energy distribution function (EEDF). In this work, we used the three-level transition involving the $3d4F^{7/2} - 4p4D^{5/2}$ states, with excitation at 668.614 nm and fluorescence emission at 442.724 nm, as this transition yielded the highest fluorescence intensity among the transitions that is accessible with the used TDL.

For time-averaged PLIF measurements (**Section 4.1**), the laser beam is modulated using a mechanical chopper to maintain consistency with a conventional LIF setup [45], ensuring proper validation of the developed system. For phase-resolved measurements (**Section 4.2**), the mechanical chopper is removed from the setup for operational convenience. The laser beam is delivered to the chamber via a fiber coupler and a multimode fiber. Upon exiting the fiber, the beam enters the shaping optics, which include a large area beam collimator that produces a circular Gaussian beam of approximately 0.5-inch diameter. Two parallel razor blades are used to trim the beam edges and shape it into a semi-planar laser sheet with a maximum power of 20 mW. The resulting laser sheet is approximately 12 mm wide and 1 mm thick. Note that a set of cylindrical lenses could also be used for beam shaping. The laser sheet is aligned to propagate in the (r–z) plane in the downward z direction, enabling Doppler shifts due to the radial motion of the metastable argon ions.

The collection branch consists of a set of lens objectives, defining a field of view (FOV) of 15 cm × 15 cm. A 441.6 nm/3 nm bandpass filter (Andover 442FS03-12.5) is installed in front of an ICCD camera (PI-MAX 4:1024i) as part of the collection optics. Details regarding the operating parameters of the ICCD camera are described in **Section 3.2**.

Fig. 2(b) shows an image acquired from the collection system with the laser tuned to the central wavelength of 668.6138 nm. Since the commercial bandpass filter used has a relatively broad bandwidth, it cannot fully eliminate stray light such as spontaneous plasma emission. As a result, the recorded fluorescence pattern is buried under background signals. To extract the fluorescence pattern, a background image, captured with the laser blocked from entering the plasma, is subtracted from the raw image. An example of the resulting fluorescence pattern is shown in Fig. 2(c). This procedure assumes that the background plasma emission remains sufficiently stable over time, such that its drift or fluctuation is smaller than the laser-induced fluorescence during the wavelength scan. While this assumption is critical for plasmas that experience drift, it is less restrictive than what is required for the conventional PMT-based method. Since the ICCD-based system significantly reduces the total measurement time across multiple spatial positions, it is more tolerant of long-timescale plasma drifts. Further discussion on this assumption can be found on the Appendix A.

For time-averaged measurements, the fluorescence extraction procedure described above was repeated while tuning the laser wavelength around the central value of $\lambda_0 = 668.6138$ nm. Each pixel in the recorded images corresponds to a fluorescence intensity at a specific spatial location within the FOV. By applying the Doppler shift principle, IVDFs can be evaluated at each pixel, resulting in spatially resolved IVDFs.

For phase-resolved measurements, the same procedure is applied; however, the timing of ICCD imaging is synchronized with the anticathode bias voltage. The oscillating voltage signal, branched from the anticathode, is fed into the device's built-in microchannel plate (MCP) triggering circuit, enabling image acquisition at specific phase intervals of the plasma oscillation. The ICCD camera is triggered during the Collector and Repeller regimes, with an exposure time set to one-tenth of the oscillation period ($\tau/10$), to demonstrate phase-resolved LIF measurements of spatially resolved IVDFs. The upper limit of the temporal resolution is determined by the hardware specifications. For instance, the MCP triggering circuit in the ICCD camera used here supports repetition rates up to 8 kHz, allowing resolution of plasma oscillations below this frequency.

The spatial resolution and positional uncertainty of the measured IVDFs are governed by the physical area on the FOV corresponding to a single CCD pixel. Considering the magnification of the lens objectives and the pixel size of the ICCD camera used in our system, each pixel collects fluorescence from a square region of $\Delta L \times \Delta L = 0.145\ mm \times 0.145\ mm$ in the (r–z) focal plane. To suppress speckle noise caused by dark current or cosmic rays, fluorescence at each pixel is averaged with its surrounding square neighborhood, yielding an effective spatial resolution of $3\Delta L$. Additional effects such as the Nyquist limit [48] and the diffraction limit [49] can also be considered, but their impact is negligible for the system configuration and wavelength range used in this study.

The developed 2D ICCD-based PLIF system is validated against a conventional 0D PMT-based LIF system [45]. The comparison results are presented in **Section 4.1**. In the PMT-based system, spatial scanning is performed by manually adjusting the focal point using a linear translation stage in 5 mm increments. The positional uncertainty is approximately 2.5 mm, as the measured positions are manually shifted post-acquisition to align with the optical focus of the ICCD-based system. The radial center, r = 0 mm, is defined as the location with the maximum fluorescence intensity in the ICCD image, based on the near-axisymmetry of the plasma. The spatial resolution of the PMT-based LIF system is determined by the iris aperture and is set to 0.9 mm for this experiment. Table 2 summarizes the spatial uncertainties of both the PMT-based and ICCD-based PLIF systems used for cross-validation.

Table 2. Measurement uncertainties of PMT-based and ICCD-based PLIF system.

|  | **PMT-based LIF system** | **ICCD-based PLIF system** |
|---|---|---|
| Position uncertainty | $\pm 2.5\ mm$ | $\pm 0.435$ mm |
| Spatial resolution | $\pm 0.9\ mm$ | $\pm 0.435$ mm |
| Velocity uncertainty | $\pm 89$ m/s ||
| Temperature uncertainty | $0.007 eV$ ||

The signatures of ion dynamics in this plasma are further analyzed by fitting the measured IVDFs to a shifted Maxwellian (Gaussian) ion velocity distribution function,

$$f(v) = A_0 \exp\left(-\frac{M_i(v-v_0)^2}{2eT_i}\right) + d \quad (1)$$

where $A_0$ is the peak intensity, $M_i$ is the ion mass, e is the elementary charge, $v_0$ is the most probable velocity which also corresponds to the mean velocity for this distribution, $T_i$ is the ion temperature in electronvolts (eV), and d is an offset constant. While multiple-Gaussian fitting functions can be employed, particularly under certain operating conditions of this plasma device [45], a single-Gaussian assumption is sufficient for the experimental scenarios discussed here. The fitting procedure minimizes the chi-square ($\chi^2$) value using the standard deviation of the measured data as the weighting input, ensuring that the resulting fit emphasizes statistically reliable data points. This fitting analysis is performed across all spatial regions of the fluorescence image to reconstruct two-dimensional maps of ion dynamic parameters.

## 3.2. ICCD operating condition

The acquired images and the extracted fluorescence signals may exhibit varying levels of noise, depending not only on the metastable ion density, which is proportional to the fluorescence intensity for a given input laser power, but also on the operating conditions of the ICCD camera. Parameters such as exposure time, number of accumulations, intensifier gain, and the number of frames used for averaging can all significantly affect image quality. Consequently, the accuracy and clarity of the reconstructed IVDFs are directly influenced by the quality of the recorded images and the corresponding extracted fluorescence signals.

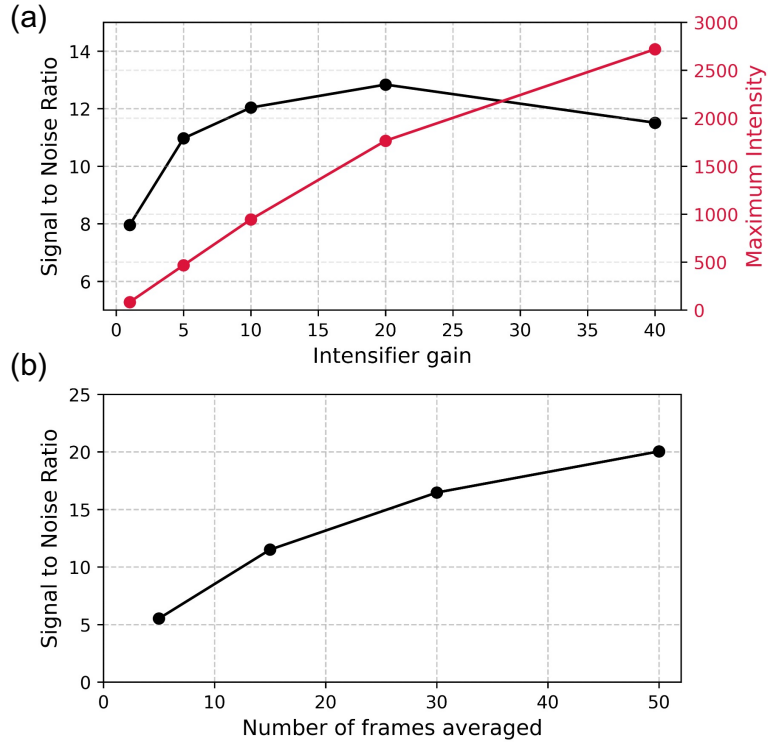

Figure 3. Effect of two ICCD acquisition parameters on image quality, quantified by the signal-to-noise ratio (SNR). The signal is defined as the maximum fluorescence intensity in the image, while noise is defined as the standard deviation of the signal across three repeated acquisitions. (a) Effect of intensifier gain, with the number of frames averaged fixed at 15. Maximum fluorescence intensity at each gain setting is also shown in red. (b) Effect of the number of frames averaged, with the intensifier gain fixed at 40. All parameter scans are conducted under the reference plasma operating condition, with the incident laser wavelength set to $\lambda = 668.6138\ nm$.

The quality of images obtained under different ICCD operating conditions was evaluated by repeating the measurement procedure three times and calculating the SNR of the fluorescence patterns. In this analysis, the signal is defined as the average fluorescence intensity at the brightest pixel, and the noise is the standard deviation of that intensity across three repeated measurements. Figure 3 presents the effects of two key ICCD parameters, intensifier gain, and the number of frames averaged, on the resulting SNR.

In Fig. 3(a), the SNR initially increases as the intensifier gain is raised, but then it decreases at higher gains due to the nonlinear amplification behavior of the intensifier. At low gain levels, both signal and noise are amplified at a similar rate. At higher gain settings, the signal amplification begins to saturate, while the noise continues to grow. This leads to a decline in SNR. This behavior suggests that using higher intensifier gain helps to highlight the peak of the IVDFs, which is useful for identifying the most probable ion velocity. However, it is less effective for accurately capturing the broadening of the distribution or ion temperature, due to increased noise.

Fig. 3(b) shows the effect of the number of frames averaged. As more frames are averaged, the noise decreases, and the image quality improves. This behavior aligns with the central limit theorem, which predicts that random noise will decrease in proportion to the inverse square root of the number of samples, assuming a normal distribution. The SNR trend observed in the data qualitatively follows a square-root dependence, which supports this explanation.

For time-averaged measurements in Scenario (A), the ICCD intensifier gain was set to 40 to better resolve the peak of the IVDFs. Images were obtained by averaging 15 frames, each frame captured with a 5 ms exposure accumulated 200 times. The total time required to acquire spatially resolved IVDFs across 40 velocity points was approximately 8 minutes using the ICCD-based system. Notably, this duration is comparable to the time needed by the PMT-based system to measure the IVDF at a single spatial point. For phase-resolved measurements in Scenario (B), the exposure time was shortened to 0.2 ms, corresponding to one-tenth of the oscillation period, in order to distinguish the measured data at different phases of the plasma oscillation.

## 3.3. Image denoising methods

As shown in the previous section, increasing the number of frames is an effective and straightforward way to improve image quality and obtain more reliable IVDFs. However, this approach comes with the trade-off of increased experimental time. To address this limitation, it is worth considering noise filtering techniques instead of relying solely on frame accumulation.

Traditional plasma diagnostics that utilize camera-based imaging often apply standard spatial filtering methods, such as median or Gaussian filters. These methods reduce noise by smoothing the image across the spatial domain, but this comes at the cost of degraded spatial resolution. As a result, such techniques may not be suitable for our application, which requires accurate measurement of spatially resolved IVDFs. Since the effectiveness of denoising methods depends heavily on the structural characteristics of the image, our goal is to identify the most effective approach for processing fluorescence patterns. To this end, we quantitatively compare the performance of several image filtering techniques [50].

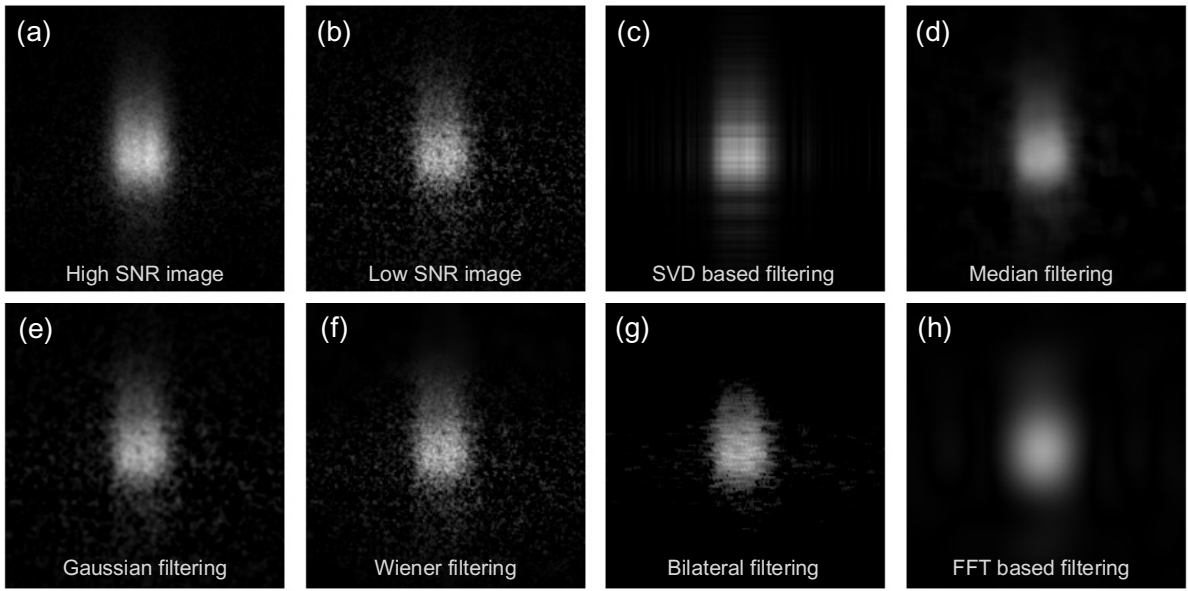

Figure 4. Fluorescence images under different ICCD operating conditions and results of various denoising methods applied to a low-SNR image. Images are cropped to zoom into the fluorescence patterns. (a) High-SNR image obtained with 50 frames averaged. (b) Low-SNR image obtained with 5 frames averaged. Low-SNR image denoised using the following image processing techniques: (c) SVD-based filtering, (d) Median filtering, (e) Gaussian filtering, (f) Wiener filtering, (g) Bilateral filtering, and (h) FFT-based filtering.

While operating the plasma under Scenario (A) with the laser set to the central wavelength, fluorescence images were acquired using two different ICCD configurations: one averaging over 50 frames (Fig. 4(a), high-SNR setup) and the other averaging over 5 frames (Fig. 4(b), low-SNR setup). As expected, the high-SNR configuration produced a clearer image, though it required ten times longer acquisition time. The objective of this analysis is to identify an effective image processing method that can denoise the low-SNR image in a way that closely approximates the high-SNR result, which is treated as the ground truth in this evaluation.

Six commonly used denoising methods were tested: Singular Value Decomposition (SVD)-based filtering, median filtering, Gaussian filtering, Wiener filtering, bilateral filtering, and Fast Fourier Transform (FFT)-based low-pass filtering. Each method was applied to the low-SNR image, and the denoised results were assessed using two quantitative metrics: Peak Signal-to-Noise Ratio (PSNR) and Structural Similarity Index Measure (SSIM). Further details on each filtering technique and the calculation of quality metrics are provided in Appendix B as well as [50]. The denoised images are presented in Fig. 4(c)–(h), and the corresponding image quality metrics are summarized in Table 3.

Table 3. PSNR and SSIM for each image denoising method applied to the low-SNR fluorescence image. The high-SNR image is used as the reference for evaluation.

|  | Low SNR | SVD | Median | Gaussian | Wiener | Bilateral | FFT-based |
|---|---|---|---|---|---|---|---|
| **PNSR** | 33.7 dB | 41.29 dB | 40.86 dB | 38.03 dB | 35.52 dB | 26.2 dB | 34.92 dB |
| **SSIM** | 0.68 | 0.84 | 0.82 | 0.74 | 0.79 | 0.18 | 0.82 |

Among the tested methods, SVD-based filtering and median filtering produced higher PSNR and SSIM values compared to the others, indicating superior denoising performance. To further evaluate their effectiveness, both methods were applied to a set of fluorescence images obtained through laser wavelength scans, as detailed in Appendix B. Overall, SVD-based filtering consistently achieved better PSNR and SSIM values, unless the median filter was applied with a larger spatial window. However, increasing the window size in median filtering results in greater image blurring, which degrades the spatial resolution of the diagnostics. This trade-off makes SVD-based filtering a more favorable choice for preserving both image quality and diagnostic resolution.

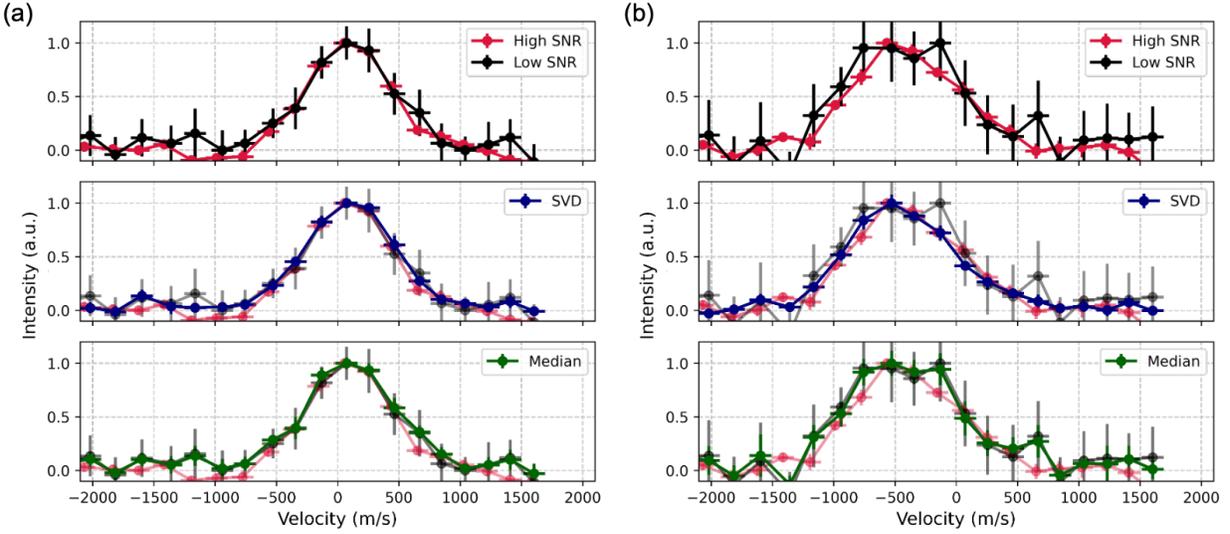

Figure 5. IVDFs at two locations: (a) r = 0 mm and (b) r = -15 mm. The IVDFs are reconstructed from images acquired using the high-SNR ICCD setup (red) and the low-SNR ICCD setup (black). Denoised results from two filtering methods are also shown: SVD-based filtering retaining the most significant component (blue), and median filtering with a window size of [51 x 51] (green).

The effectiveness of each filtering method becomes more apparent when the denoised images are converted into IVDFs. Fig. 5 presents IVDFs at two different spatial locations, reconstructed from images acquired using different ICCD setups and processed with various denoising methods. Additional IVDFs at other positions are provided in Appendix B. At the center of the plasma, where the fluorescence signal is strong due to a higher density of metastable species, both filtering methods yield similar results. In this region, the signal-to-noise ratio is inherently high, and thus the impact of denoising is minimal.

However, at the plasma periphery, where the fluorescence is weaker and the noise level is higher, the performance of SVD-based filtering becomes more significant. This method is effective at isolating the dominant structural features of the image from noise, allowing it to recover the underlying fluorescence pattern even under low signal conditions. As a result, the IVDFs reconstructed from SVD-denoised images more closely follow the ground truth, high-SNR curve, while also exhibiting reduced noise levels. In contrast, median filtering, while useful for general noise suppression, also blurs the finer features of the fluorescence pattern. Hence, it fails to preserve the correct shape of the distribution. This leads to a loss of detail in the reconstructed

IVDF, especially at the 'wing' region where SNR is low but contains the higher moment characteristics of IVDFs such as ion temperature.

Therefore, beyond the PSNR and SSIM metrics provided earlier, the reconstructed IVDFs further demonstrate that SVD-based filtering is more effective than median filtering for extracting fluorescence signals from noisy images. This advantage can be attributed to the selective denoising capability of SVD-based filtering, which isolates and retains the most relevant components for image reconstruction [51,52]. In addition to its denoising performance, SVD-based filtering also preserves the spatial resolution of the measurements, as it avoids the blurring commonly introduced by conventional spatial filters, making it well-suited for spatially resolved IVDF measurements.

# 4. Results & Discussion

## 4.1. Time-Averaged Measurements

This section presents time-averaged measurements of IVDFs obtained using the developed ICCD-based PLIF system. The validation of this diagnostic approach is discussed through a direct comparison with a conventional PMT-based LIF system. In addition, the spatially resolved characteristics of the 2D IVDFs are fitted to (1) and analyzed based on key parameters: peak intensity, most probable velocity, and ion temperature.

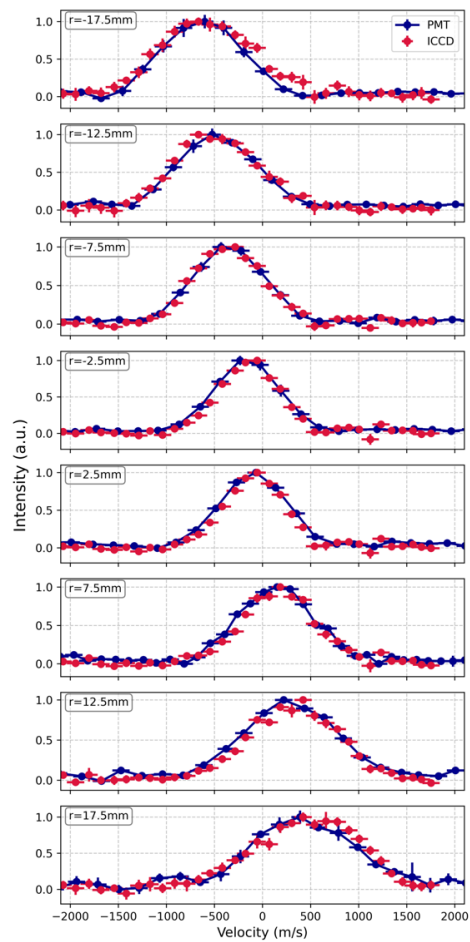

Figure 6. Comparison of IVDFs at different spatial locations, measured using PMT-based LIF (blue) and ICCD-based PLIF (red). The x-axis indicates ion velocity (m/s), and the y-axis shows fluorescence intensity in arbitrary units, normalized by the maximum value of each IVDF to enable convenient comparison across positions.

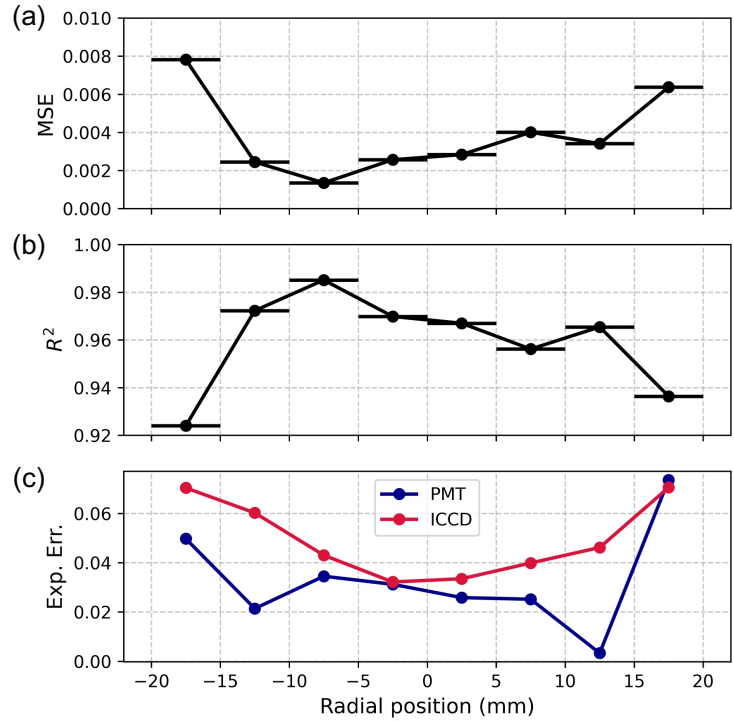

Figure 7. Quantitative comparison of measured IVDFs from PMT-based LIF and ICCD-based PLIF at different positions, using (a) mean squared error (MSE), (b) coefficient of determination ($R^2$), and (c) the average experimental error bars of the IVDFs measured by each system at each position.

Fig. 6 shows the IVDFs measured in different radial positions of this plasma in the reference operating condition. In each plot, the x-axis represents ion velocity (m/s), and the y-axis shows the fluorescence intensity normalized by the maximum value of the IVDF at each position to facilitate comparison. Error bars on the fluorescence intensity indicate the standard deviation from three repeated measurements. The ion velocity error bars represent a combination of the standard deviation from three measurements and the wavemeter uncertainty, which is approximately ± 90 m/s. The radial position corresponding to each IVDF is labeled in each subplot. Figure 7 shows the differences between IVDFs measured by each diagnostic, quantified using the mean squared error (MSE) and the coefficient of determination ($R^2$). The discrepancies between the two measurements are more vivid at the plasma periphery than at the radial center, where the metastable density, and consequently, the fluorescence intensity and SNR are lower. As shown in Fig. 7(c), both diagnostics exhibit larger experimental error bars at the periphery, contributing to the observed deviations. Nevertheless, the MSE remains below 1% across all positions, and R² values exceed 0.9, indicating excellent agreement between the developed ICCD-based PLIF system and the conventional PMT-based LIF system.

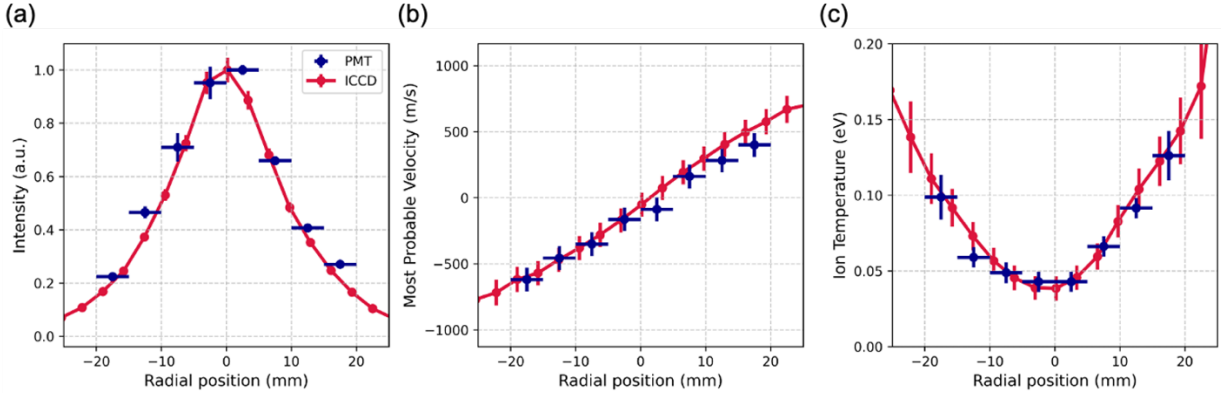

Figure 8. Radial distribution of (a) Peak fluorescence intensity, (b) Most probable velocity (m/s) and (c) Ion temperature (eV) measured by PMT-based LIF (blue) and ICCD-based PLIF (red).

Fig. 8 shows the 1D radial distribution of the peak intensity ($A_0$), most probable velocity ($v_0$), and the ion temperature ($T_i$) are extracted from the above 2D map, along z = 0 mm. Same fitting methods are repeated to IVDFs measured by PMT-based LIF method, and overplotted for comparison. The y error bars for each fitting parameters show the 95% confidence interval of the $\chi^2$ fitting, indicating the statistical stability of the fit. The x error bars indicate the uncertainty of the spatial positions, as shown in Table 1. The fitted parameter from both methods agrees with each other as well as IVDFs, also supporting a successful cross-validation of the developed PLIF system against conventional measurements.

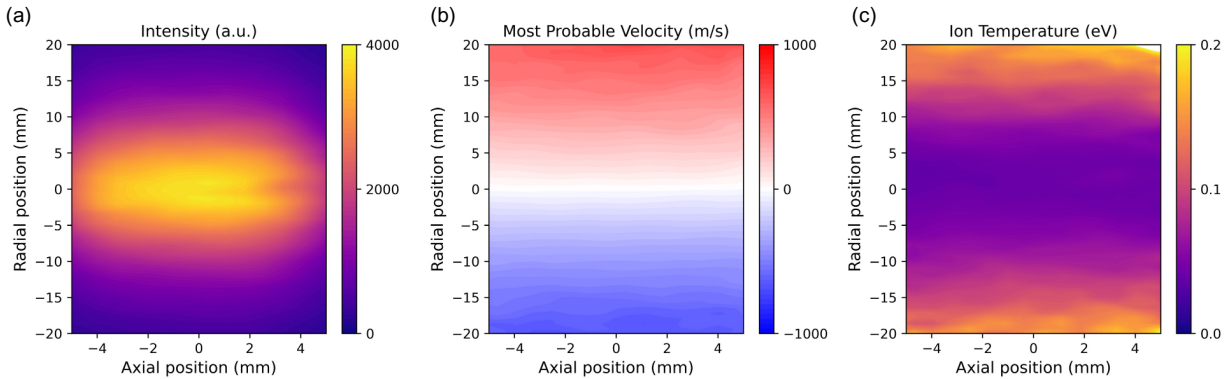

Figure 9. Experimentally obtained two-dimensional characteristics of ion dynamics, analyzed through fitting of the IVDFs. (a) Peak fluorescence intensity, (b) Most probable ion velocity (m/s), (c) Ion temperature (eV).

The spatially resolved characteristics of 2D IVDFs are shown in Fig. 9. In each contour plot, the x-axis and y-axis represent axial and radial positions, respectively. Due to the near-axisymmetry and axial uniformity of the plasma, the fitted parameter maps exhibit radial symmetry and homogeneity along the z-direction. For visualization, only the spatial averages of the fitted parameters are shown in Fig. 9. It is important to note that fitting uncertainties increase toward the periphery of the plasma, where fluorescence intensity is weaker, and the SNR is lower. Consequently, some apparent spatial variations, such as the wavy structures observed in the axial direction of the ion temperature map in Fig. 9(c) are likely artifacts within the margin of fitting error rather than physical features.

Fig. 9(a) shows the spatial map of the peak fluorescence intensity ($A_0$) at each position, which depends on both the local laser power and the density of ion metastables excited by the laser. The non-uniformity observed along the axial (z) direction is primarily attributed to the spatial variation in the laser sheet's power distribution. The radial decrease in peak intensity reflects the reduction in ion metastable density, which is expected due to plasma diffusion.

Fig. 9(b) presents the map of the most probable velocity ($v_0$). This quantity represents the overall ion flow direction and magnitude. The map reveals outward ion streaming from the radial center, consistent with ambipolar diffusion processes that maintain charge neutrality. This behavior aligns with predictions from modeling studies of this plasma source [44].

Fig. 9(c) displays the ion temperature ($T_i$) distribution. Near the radial center, the ion temperature is approximately 0.045 eV. Toward the plasma periphery, the ions become significantly hotter, reaching temperatures up to ~0.15 eV at r = 20 mm, which is more than three times higher than at the center. Particle-in-cell (PIC) simulation studies conducted on this plasma source under conditions similar to the experiment scenario (A) have proposed a heating mechanism associated with the rotating spoke instability [53,54]. Our observation provides experimental evidence of these warm ions at the periphery, which may originate from the instability-driven heating mechanism. A more detailed experimental investigation of the interplay between the rotating spoke and ion heating is reserved for future work.

## 4.2. Phase-Resolved Measurements

In this section, the applicability of this ICCD-based PLIF system on phase-resolved measurements is discussed, demonstrating IVDF measurements in plasma operating scenario (B) where plasma is driven with an external oscillation applied at anticathode.

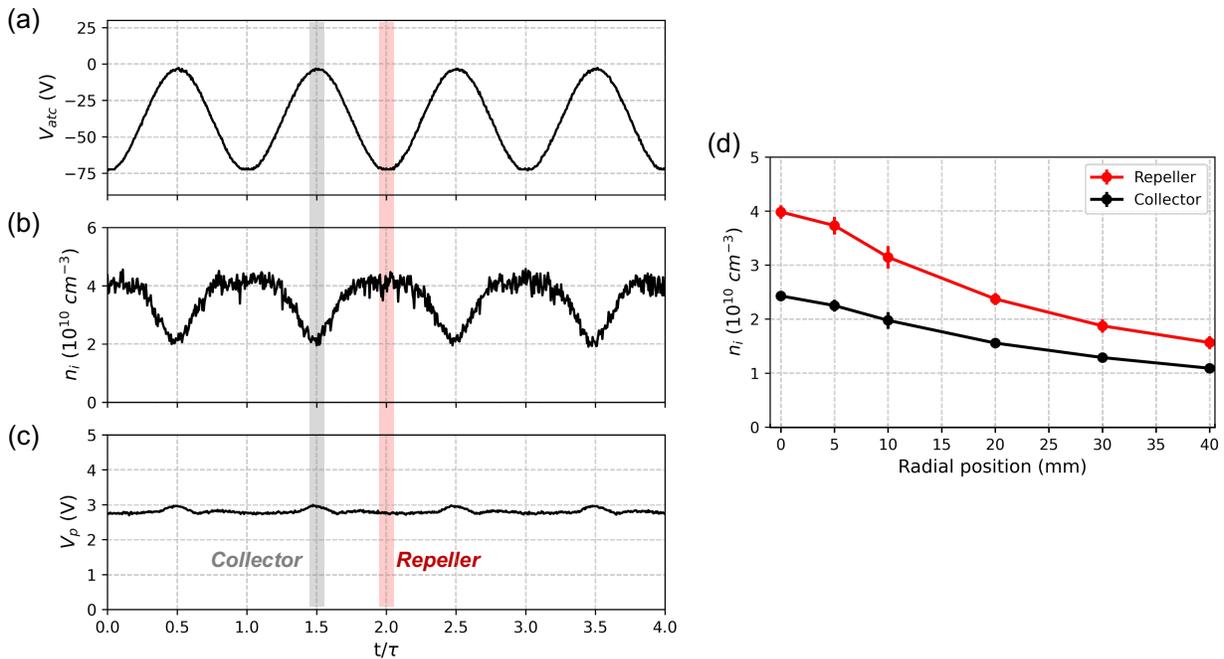

Figure 10. Time evolution of key plasma parameters at the radial center, synchronized with the modulation of the anticathode potential. (a) Anticathode potential, (b) Ion density, (c) Plasma potential. (d) Radial profiles of ion density during Collector mode and Repeller mode measured by an ion probe.

Fig. 10(a)-(c) shows the time evolution of key plasma parameters in response to the oscillating anticathode potential. The plasma oscillates between two distinct operating conditions, referred to as Collector and Repeller mode [44,46]. The x-axis represents time normalized by the oscillation period of 2 ms. The ICCD camera is triggered at two specific phases corresponding to the Collector and Repeller modes, as indicated by the gray and red boxes, respectively. The width of each box denotes the exposure time $\tau/10$ following the trigger signal.

The radial profiles of ion density, measured using an ion probe, are presented in Fig. 10(d). The ion density is higher in the Repeller mode compared to the Collector mode, as reduced axial loss of the electron beam leads to increased ionization. Notably, the ion density at the center is approximately 1.6 times greater in Repeller mode than in Collector mode. The observed variation in ion density provides a valuable test environment for evaluating the performance of the phase-resolved measurements of the developed PLIF system. Previous study provides that the electron temperature and electric potential structure remains consistent between the two modes [46]. Hence, the fluorescence intensity at each mode is expected to scale proportionally with the ion density.

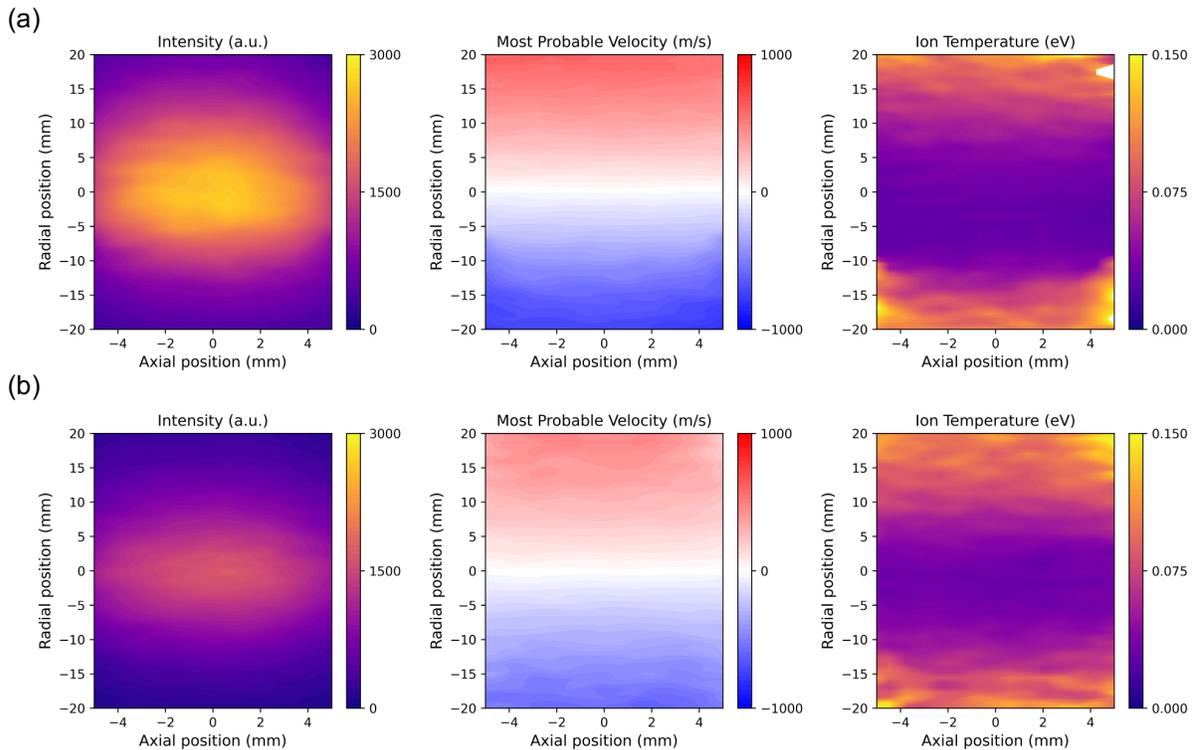

Figure 11. Experimentally obtained two-dimensional characteristics of ion dynamics, analyzed through fitting of IVDFs at two different ICCD triggering phases: (a) *Repeller* mode, (b) *Collector* mode. Each row displays spatial maps of peak fluorescence intensity, most probable ion velocity (m/s), and ion temperature (eV), extracted through fitting of the IVDFs in each mode.

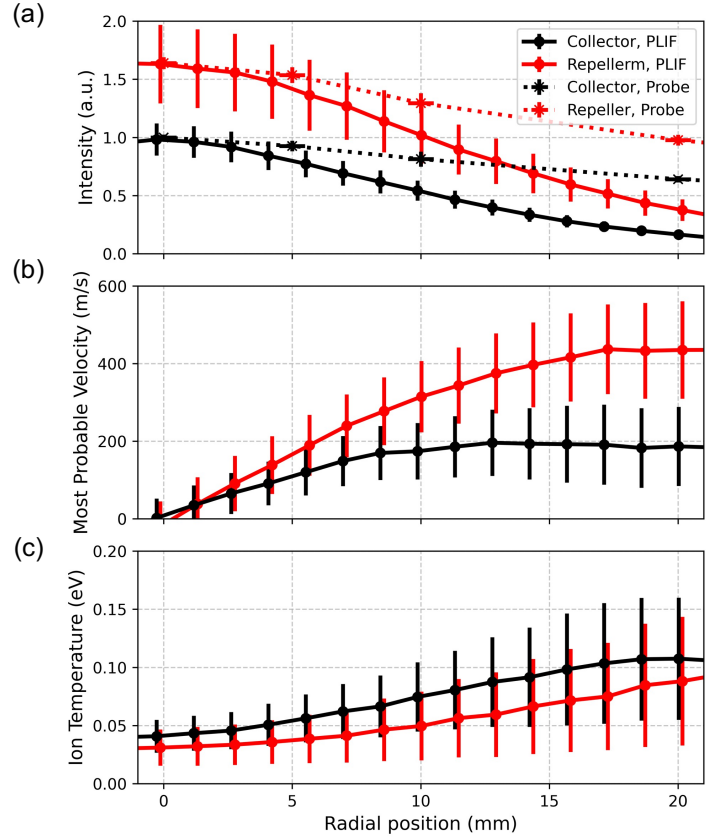

Figure 12. Radial distribution of ion dynamic parameters in *Repeller* mode (red) and *Collector* mode (black): (a) Peak fluorescence intensity, with the normalized ion density profile overlaid as a dotted line; (b) Most probable ion velocity (m/s); (c) Ion temperature (eV).

The IVDFs obtained from phase-resolved PLIF measurements are fitted using a shifted Maxwellian distribution function. The resulting 2D maps of IVDF signatures are presented in Fig. 11. These phase-resolved 2D maps exhibit axisymmetry, radial ion outflux and ion heating near the plasma periphery, similar to the time-averaged measurements. The corresponding 1D radial profiles at $z = 0$ mm are shown in Fig. 12. Fig. 12(a) displays the peak fluorescence intensity profiles for each mode, normalized to the peak fluorescence at the radial center in Collector mode. For comparison, the ion density profiles in each mode, also normalized to the central density in Collector mode, are overlaid. The fluorescence intensity at the radial center in Repeller mode is approximately 1.6 times higher than in Collector mode, consistent with ion density measurements obtained via probe diagnostics. The radial distribution of peak fluorescence intensity in each mode has a same trend to the ion density profile, as expected, but exhibits a steeper gradient. This difference in slope is attributed to the center-peaked electron temperature distribution [46], which enhances fluorescence by exciting more metastable states from the same ion density. The alignment of these fluorescence profiles with probe-measured density profiles supports the successful implementation of phase-resolved PLIF measurements.

Differences in IVDF characteristics between the two modes further validate the system's phase-resolving capability. For instance, the most probable ion velocity is higher in *Repeller* mode, likely due to enhanced plasma diffusion driven by a steeper density gradient and a more pronounced center peak. The ion temperature profile reveals that ions are approximately twice as hot at the periphery compared to the center. This trend is similar to what was observed in Scenario (A), though to a lesser extent. Specifically, the ratio of ion temperature at r = 20 mm to the center temperature is around 2 in both modes of Scenario (B), whereas it was approximately 3 in Scenario (A), as discussed in **Section 4.1**. This observation further supports the proposed ion heating mechanism associated with the rotating spoke instability. The higher pressure in Scenario (B) suppresses spoke activity [47], which may in turn reduce the level of ion heating observed at the plasma edge.

# 5. Conclusion

In this work, we developed and tested a planar laser-induced fluorescence (PLIF) system for space- and phase-resolved ion velocity distribution function (IVDF) measurements, using an electron-beam generated E × B plasma source as an experimental test bench. The influence of ICCD camera operating parameters and image denoising techniques was systematically evaluated, providing a practical framework for selecting optimal acquisition settings and filtering methods. The developed ICCD-based PLIF system was successfully validated against a conventional PMT-based LIF system, which was mechanically scanned over different spatial locations. Notably, the ICCD-based PLIF significantly reduces total measurement time by capturing the entire spatial map in the time required by the conventional method to measure a single point.

The developed PLIF system captured key signatures of ion dynamics in this plasma source. Time-averaged measurements captured radial ion outflow driven by plasma diffusion, as well as anomalous ion heating at the plasma periphery. The detection of warm ions offers experimental support for previous theoretical studies, which proposed a link between ion heating and rotating spoke instabilities in this system. Phase-resolved measurements captured the dynamic oscillation between *Repeller* and *Collector* modes. The close agreement between the fluorescence intensity maps and probe-diagnosed ion density profiles, along with distinct ion signatures at different phases, further confirms the successful implementation of the space and phase-resolved IVDF measurement.

Overall, the developed PLIF system enables time-efficient, space and phase-resolved IVDF measurements, paving the way for future advances in plasma physics questions. In particular, its phase-resolving capability offers a powerful tool for investigating ion dynamics associated with plasma instabilities, such as rotating spokes, and for uncovering the complex interplay between instabilities and particle kinetics in low-temperature plasmas. This work also highlights opportunities for further diagnostic development, including the integration of advanced imaging devices such as EMICCDs, which could improve temporal resolution and expand the system's phase-resolving capability.

# APPENDIX

## A. Discussion on the background subtraction

The proposed ICCD-based PLIF system is more tolerant on plasma drift on longer time scale compared to the conventional LIF system, as mentioned in **Section 3.2**. However, it is worth to discuss further about the criticality of this assumption on the further possible application of this system to different plasma sources. We assume that the background emission from the plasma remains sufficiently stable over time, such that its drift or fluctuation is smaller than the laser-induced fluorescence. This allows a single background image to be used for all fluorescence images acquired at different laser wavelengths and at different times. If, however, the background emission fluctuates rapidly and more strongly than the fluorescence signal, inaccurate background subtraction may lead to failure in proper extraction of the fluorescence.

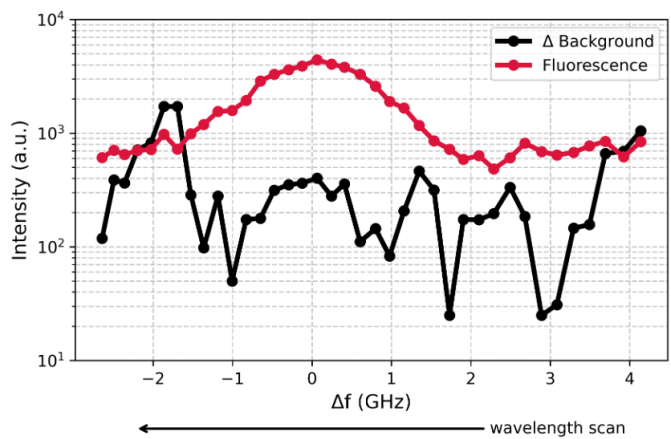

Figure 13. Log-scale plot of the intensity drift in the background image and the maximum intensity of the extracted fluorescence image during the laser wavelength (frequency) scan.

To evaluate the reliability of the above assumption in our plasma source, additional measurements were conducted under scenario (A). Using a mechanical chopper as the trigger for the ICCD camera, one image with laser incidence and one background image without laser incidence were recorded at each laser wavelength during the scan, separated by a 10 ms interval. This short delay allows the two images to be considered as captured at the same macroscopic time scale. Fig. 13 presents the deviation of the maximum intensity in the background images and the maximum intensity of the extracted fluorescence images throughout the scan. The scan sequence is indicated by an arrow, and the entire process took approximately 8 minutes, comparable to the duration required for a single-point measurement using the conventional LIF method. The results show that the variation in background signal is orders of magnitude smaller than the fluorescence signal within the wavelength range where the peak and broadening of the IVDFs are primarily resolved. However, at the tail of the scan (e.g., around $\Delta f = -2$ GHz), background deviations can exceed the fluorescence signal if the latter is weak. Therefore, for broader implementation of the proposed PLIF system in different plasma sources, it is essential to quantify background drift or fluctuation relative to the fluorescence signal to assess the system's feasibility.

B. Details on the image denoising methods

In this section, we detail the analysis of image denoising results introduced in **Section 3.3.** Six denoising methods are applied to the low-SNR image (Fig. 4(b)) and compared against the high-SNR reference image (Fig. 4(a)) using two quantitative metrics: peak signal-to-noise ratio (PSNR) and structural similarity index measure (SSIM) [50].

PSNR evaluates the fidelity of a denoised image by calculating the ratio between the maximum possible pixel intensity and the mean squared error (MSE) between the denoised image and the reference. Higher PSNR values indicate better denoising performance, as they correspond to a smaller deviation from the ground truth. However, PSNR is based solely on pixel-wise intensity differences and may not fully capture perceived image quality.

SSIM, in contrast, assesses the perceptual similarity between images by incorporating structural information, luminance, and contrast. It produces values ranging from -1 to 1, with 1 indicating perfect structural similarity. Because it captures texture and structural content, SSIM is more aligned with human visual perception and is therefore often more appropriate for evaluating denoised image quality.

The six denoising methods evaluated are: Singular Value Decomposition (SVD)-based filtering, median filtering, Gaussian filtering, Wiener filtering, bilateral filtering, and Fast Fourier Transform (FFT)-based filtering. The working principle and the filtering parameters of each filter is introduced below.

*SVD-based filtering* works by decomposing the image into singular values and corresponding singular vectors, enabling noise suppression through the selective retention of dominant components. This method is effective in enhancing image quality by preserving essential structures while removing noise. The number of retained components is varied from 1 to 3 to find the optimal filtering parameter. For instance, "SVD, 2nd" in this analysis refers to retaining only the first and second most significant components, while discarding the rest assumed to contain noise.

*Median filtering* is a nonlinear filtering technique that replaces each pixel value with the median of the intensities in its neighborhood window [N x N], effectively removing impulsive noise. It is particularly effective in suppressing salt-and-pepper noise due to its ability to reject outliers. The window size is varied from 11 to 51 to find the optimal filtering parameter.

*Gaussian filtering* applies a convolution operation with a Gaussian kernel, providing smooth image results by reducing high-frequency noise components. It achieves noise suppression through weighted averaging, where pixels closer to the center have higher weights, ensuring minimal distortion of the image details. The width of the Gaussian kernel $\sigma$ is varied from $\sigma = 1$ to 3.

*Wiener filtering* is a linear filtering technique that minimizes the mean square error between the estimated and true signal, under the assumption of a stationary signal and noise process. It adapts to local image variance, providing stronger noise suppression in high-SNR regions while preserving important image features. The primary tuning parameter is the size of the local neighborhood window [N x N], which determines the area over which local statistics (mean and variance) are computed for adaptive filtering. The size of the local window is scanned over the range 11 to 51.

*Bilateral filtering* preserves edges while reducing noise by considering both spatial proximity and intensity similarity when averaging pixel values. It applies a weighted average, where the weights are determined by two Gaussian functions: one based on geometric closeness (spatial domain) and the other on photometric similarity (intensity domain). The degree of smoothing is controlled by two parameters, $\sigma_s$ for spatial distance and $\sigma_r$ for intensity difference. Smaller $\sigma$ values result in stronger edge preservation, while larger values lead to more aggressive smoothing. It is widely used for denoising and edge-preserving smoothing in computer vision and image processing tasks. In this study, the spatial variance $\sigma_s$ is scanned over the range $\sigma_s = 1$ to 3.

*FFT based filtering* operates in the frequency domain by transforming the image, filtering out high-frequency components associated with noise, and then performing an inverse transform to reconstruct the denoised image. This approach allows selective attenuation of noise frequencies while preserving important structural components. FFT-based filtering is particularly effective for periodic noise removal and signal processing applications. Here, the high-frequency cutoff is varied from 0.01 to 0.5 to find the optimal setting.

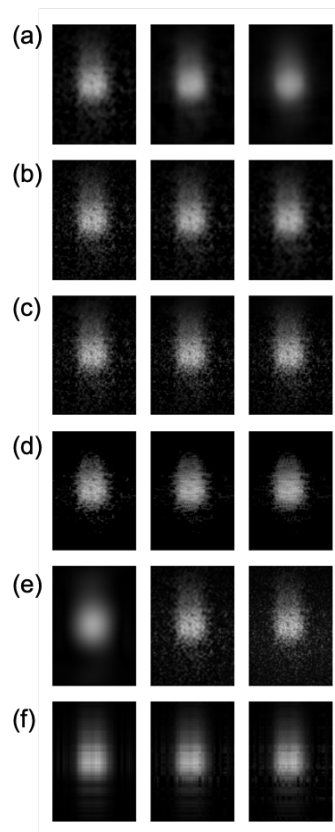

Figure 14. Low-SNR denoising results of the fluorescence image at the laser frequency of $\Delta f = 0$, using various filtering parameters. Rows correspond to different methods, and columns within each row illustrate changes in image quality when different denoising parameters are applied. (a) Median filtering with window sizes of 11, 25, and 51. (b) Gaussian filtering with $\sigma = 1$, 2, and 3. (c) Wiener filtering with window sizes of 11, 25, and 51. (d) Bilateral filtering with $\sigma = 1$, 2, and 3. (e) FFT-based low-pass filtering with cutoffs of 0.01, 0.1, and 0.5. (f) SVD-based filtering retaining up to ranks 1, 2, and 3.

The resulting denoised images with different parameters are shown in Fig. 14, while the result with the highest PSNR and SSIM for each method is played in Fig. 4. As summarized in **Section 3.3**, SVD-based filtering gave the best denoising results.

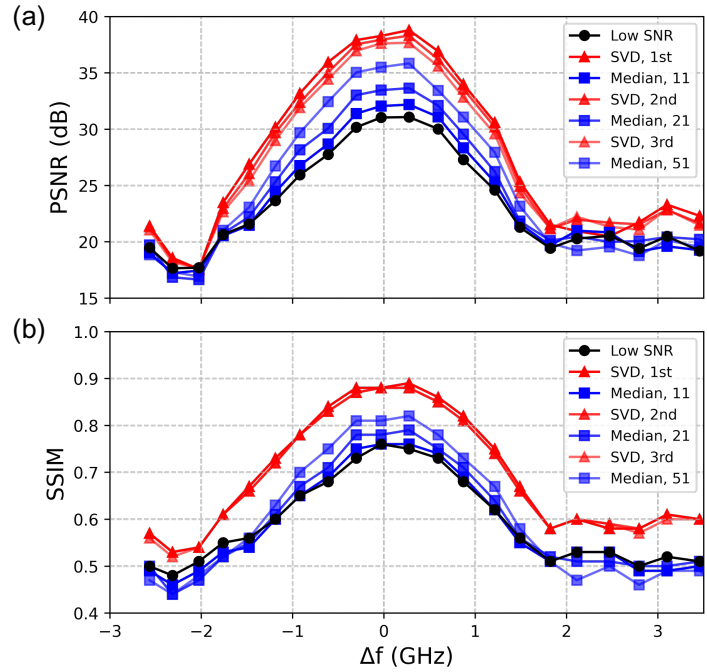

Figure 15. Quantitative evaluation of denoising performance across laser wavelength (frequency) scan $\Delta f$ using (a) PSNR and (b) SSIM metrics. Results are shown for fluorescence images reconstructed using SVD-based filtering and Median filtering. The influence of denoising parameters is illustrated by curves of the same color but varying transparency for each method. Labels such as "SVD, 2nd," indicate SVD-based filtering including components up to the 2nd significance. Labels such as "Median, 21" refer to median filtering applied with a [21 x 21] square window.

Fig. 15 shows the PSNR and SSIM in different laser wavelength, which gives differently patterned fluorescence images. Overall, the SVD-based filtering gives better PSNR and SSIM unless median filter uses a larger window of filtering so that the image is more blurred, and we lose the spatial resolution. On the other hand, the above plot shows that retaining one component from the SVD-based filtering gives the best PSNR, which indicates that the most significant (1st) component contains the fluorescence structure, while components lower than 2nd significance contains noise. Therefore, selecting only the first component in SVD-based filtering is sufficient for effective denoising across all laser wavelengths used to reconstruct the IVDFs.